\documentclass[12pt]{article}
\usepackage{epsfig}
\usepackage{amsmath}
\usepackage{amssymb}
\usepackage{latexsym}
\usepackage{graphicx}
\newcommand{\p}{\partial}

\newcommand{\sectiono}[1]{\section{#1}\setcounter{equation}{0}}
\renewcommand{\theequation}{\thesection.\arabic{equation}}

\begin{document}
{}~ \hfill\vbox{\hbox{hep-th/0406023}\hbox{MIT-CTP-3500}}\break

\vskip 3.8cm

\centerline{\large \bf Solving Witten's  SFT by Insertion of
Operators on Projectors }
\vspace*{10.0ex}

\centerline{\large \rm Haitang Yang }
\vspace*{8.0ex}
\centerline{\large \it Center for Theoretical Physics}

\centerline{\large \it Massachussetts Institute of Technology,}

\centerline{\large \it Cambridge, MA 02139, USA}
\vspace*{1.0ex}
\centerline{E-mail: hyanga@mit.edu}

\vspace*{10.0ex} \centerline{\bf Abstract}
\bigskip
\smallskip
Following Okawa, we insert operators at the boundary of regulated
star algebra projectors to construct the leading order tachyon
vacuum solution of open string field theory. We also calculate the
energy density of the solution and the ratio between the kinetic
and the cubic terms. A universal relationship between these two
quantities is found. We show that for any twist invariant
projector, the energy density can account for at most $68.46\%$ of
the D25-brane tension. The general results are then applied to
regulated slivers and butterflies, and the next-to-leading order
solution for regulated sliver states is constructed.

\baselineskip=16pt

\sectiono{Introduction and Summary}
The tachyon instability in open string theory comes from the
expansion of the action around the perturbative vacuum. Sen
conjectured that there is a nonperturbative vacuum where
D25-branes decay and only closed string excitations exist
\cite{9911116, 9902105}. However, no analytic solution that
represents the stable tachyon vacuum has been found so far. Level
truncation is a powerful method to construct numerical solutions
\cite{9902105}-\cite{0211012}. Very impressive results have been
obtained in \cite{0211012, 0002237}, where the D25-brane tension
was reproduced at very high accuracy. Level truncation, however,
can only give solutions satisfying the equation of motion when
contracted with the solutions themselves, as opposed to an
arbitrary state in the Fock space\footnote{Some other exact
solutions based on identity state were also constructed
\cite{Horowitz1986}-\cite{0304261}. Though these solutions do
solve the equation of motion in general way, singularities arise
as one tries to calculate the energy density.}.

A new approach\footnote{Another approch based on Moyal star
algebra was proposed in \cite{0302151}.} was developed by Okawa
\cite{0311115} who inserted operators at the middle point of the
boundary of regulated butterfly states
\cite{0310264}-\cite{0202151}. This approach has some similarities
to level truncation. One inserts linear combinations of operators
with the same mass dimension order by order. The solution, in the
case of butterflies, is a series expanded with powers of
$\sqrt{1-t}$, where $t$ is the regulation parameter and  $t\to 1$
gives the exact butterflies. The solution obtained by this
approach satisfies the equation of motion when contracted with
arbitrary states in the Fock space. At the same time, finite
results are obtained when calculating the energy density.

It is of interest to see if this approach can also be applied to
other star algebra projectors. The butterfly state is just a
special case of generalized butterfly states characterized by a
parameter $\alpha$ \cite{0202151}, with $\alpha=1$ for butterfly
sate. It is natural to ask if the results depend on the parameter
$\alpha$ or not. Can we write down an analytical expression for
the energy density obtained from solution based on a general star
algebra projector?

We are only concerned with twist invariant projectors in this
paper. For this kind of projectors, the middle point of the
boundary can be defined to be the point which reaches the string
middle point as the regulator is removed. All the projectors
currently known have this property. A twist invariant projector
$|P\rangle$ can be defined by:
\begin{equation}
\langle \phi|P\rangle=\langle f\circ
\phi(0)\rangle_{\mathbb{H}_z},\hspace{5mm}z=f(\xi),
\end{equation}
where $\mathbb{H}_z$ stands for the full upper half $z$ plane and
$\langle \phi|$ is an arbitrary state in the Fock space, inserted
at the puncture $\xi=0$ in the local coordinate patch. The
ambiguity of $f(\xi)$ is removed, up to a scale, by requiring
$f(0)=0$ and $f(1)=-f(-1)$. The definition of a regulated
projector $|P_t\rangle$ is:
\begin{equation}
\langle \phi|P_t\rangle=\langle f_t\circ
\phi(0)\rangle_{\mathbb{H}_z},\hspace{5mm}z=f_t(\xi),
\end{equation}
on $z$ representation, where $t$ is the regulation parameter.
$f_t(\xi)$ is required to satisfy the conditions:
\begin{equation}
f_t(0)=0,\hspace{5mm}f_t(1)=-f_t(-1),\hspace{5mm}\lim_{t\to 0}
f_t(\xi)=f(\xi).
\end{equation}
The operators will be inserted at the boundary midpoint of the
regulated projectors. In the $z$ representation of twist invariant
regulated projectors, this point is located at infinity, so it is
convenient to map it to the origin. For this purpose, we introduce
a new $\eta$ representation. $\eta$ is obtained as $\eta=I\circ
z=-\frac{1}{z}$.

In order to calculate the leading order solution that represents
the tachyon vacuum, besides the map $\eta=I\circ f_t(\xi)$, one
also needs a map $\eta'=I\circ\tilde f_t(\xi)$, where $\tilde
f_t(\xi)$ defines  the star product of two regulated projectors in
$z$ representation:
\begin{equation}
\langle \phi|P_t*P_t\rangle=\langle\tilde
f_t\circ\phi(0)\rangle_{\mathbb{H}_z},
\end{equation}
with $\tilde f_t(0)=0$ and $\tilde f_t(1)=-\tilde f_t(-1)$. To
simplify the calculations, we will scale $\tilde f_t(\xi)$ to have
the limit $\tilde f_t(\xi)\to f(\xi)$ as $t\to 0$. Therefore, one
can construct a map $\eta'=g_t(\eta)$ with $g_t(0)>0$ which maps
two copies of the surface associated with $|P_t\rangle$ to the
glued surface associated with $|P_t*P_t\rangle$. $g_t(\eta)$ plays
the role of star product and tells us where the boundary midpoints
of the two surfaces associated with $|P_t\rangle$'s are located on
the surface associated with $|P_t*P_t\rangle$. We find the leading
order solution is:
\begin{equation}
\psi^{(0)}=-a_t|P_t(c)\rangle,
\end{equation}
where $a_t=\frac{g_t'(0)^2}{2g_t(0)}>0$ and $|P_t(c)\rangle$
denotes a general twist invariant regulated projector with a $c$
ghost insertion at the boundary middle point.

Inner products of projectors are encountered when one calculates
the energy density. We will assume the availability of a map
$h(\eta)$, which maps a single regulated projector in $\eta$
representation, after cutting the local coordinate patch and
gluing the left half string with the right half string, to a unit
disk. $h(\eta)$ is fixed by mapping the boundary midpoint to $1$
and string midpoint to the origin on the unit disk. We show that
the ratio of the kinetic term to cubic interaction is:
\begin{equation} R=\lim_{t\to 0}\frac{\langle
\psi^{(0)}|Q_B|\psi^{(0)}\rangle} {\langle
\psi^{(0)}|\psi^{(0)}*\psi^{(0)}\rangle}=-2^4\cdot
3^{-9/2}\lim_{t\to 0}\frac{b_t}{a_t},
\end{equation}
where $b_t=-ih'(0)>0$. Since both $a_t$ and $b_t$ are positive,
$R$ is negative. Ideally, $R=-1$ from the equation of motion. The
limit $\lim_{t\to 0}b_t/a_t$ is conformally invariant because the
scale freedom of $a_t$ is cancelled by that of $b_t$. The ratio of
the energy density at the solution and the D25-brane tension is:
\begin{eqnarray}
\frac{\cal E}{T_{25}}&=& 2\pi^2\Big[\frac{1}{2} \langle
\psi^{(0)}|Q_B|\psi^{(0)} \rangle+\frac{1}{3}\langle
\psi^{(0)}|\psi^{(0)}*\psi^{(0)}\rangle\Big]\nonumber\\
&=&2\pi^2 \lim_{t\to 0}\Big[-8\Big(\frac {a_t}{b_t}\Big)^2+
3^{7/2}\Big(\frac {a_t}{b_t}\Big)^3 \Big].
\end{eqnarray}
A universal relationship between $R$ and ${\cal E}/T_{25}$ is
found:
\begin{equation}
\frac{\cal E}{T_{25}}=- \frac{\pi^2}{3}\left(\frac{4}{3\sqrt
3}\right)^6\frac{3R+2}{R^3}\simeq -0.6846\times \frac{3R+2}{R^3},
\end{equation}
where $-0.6846$ is the familiar value of ${\cal E}/T_{25}$ at
level zero obtained in \cite{9912249}. The quantity ${\cal
K}(R)\equiv\frac{3R+2}{R^3} $ reaches its maximum at the optimal
value $R=-1$: ${\cal K}(-1)=1$.\footnote{In a private
communication, Okawa conjectured that at the leading order, for an
arbitrary projector, $\Big|\frac{\cal E}{T_{25}}\Big|\leq 0.6846$.
Our results confirmed this conjecture because $R$ should be close
to $-1$ for a solution.} When applied to regulated butterflies,
these equations reproduce the results of \cite{0311115}. When
applied to regulated slivers, namely wedge states, it gives:
\begin{equation}
R=-2\left(\frac{4}{3\sqrt 3}\right)^3\simeq -0.9124,\hspace{5mm}
\frac{\cal E}{T_{25}}=2\pi^2\,\frac{3^{7/2}-64}{8} \simeq -0.6645.
\end{equation}
In the calculations, two maps $h(\eta)$ and $\tilde f_t(\xi)$ are
needed. Generically, it is not difficult to calculate $h(\eta)$.
However, obtaining $\tilde f_t(\xi)$ is challenging. For
butterflies, this map \cite{0202139} takes a rather complicated
form. On the other hand, $\tilde f_t(\xi)$ for regulated slivers
is very simple. Higher order calculations are much simplified if
one starts from regulated slivers. Therefore, it is also of
interest to calculate the next to leading order results based on
slivers. We find that there is only one solution with:
\begin{equation}
R\simeq -0.9209,\hspace{5mm}\frac{\cal E}{T_{25}}=-0.81736.
\end{equation}

The organization of this paper is as follows. In section 2, we
construct the leading order solution for a general twist invariant
regulated projector and calculate the energy density. We then
apply the results to regulated sliver and butterfly states in
section 3. Section 4 is devoted to the calculation of the next to
leading order solution based on regulated slivers. In section 5,
we give our conclusions.

\sectiono{The leading order solution for general star algebra projectors}

The action of Witten's cubic string field theory is
\cite{Witten1986}:
\begin{equation}
S[\Phi]=-\frac{1}{g^2}\left[\frac{1}{2} \langle\Phi|Q_B |\Phi
\rangle +\frac{1}{3}\langle \Phi|\Phi *\Phi\rangle \right],
\label{action}
\end{equation}
where  $Q_B$ stands for the BRST operator and $g$ is the open
string coupling constant. The equation of motion of this action
takes the form:
\begin{equation}
Q_B|\Phi\rangle+|\Phi *\Phi\rangle=0.
\label{EOM}
\end{equation}
A solution $|\psi\rangle$ of this theory should satisfy the
equation of motion when contracted with an arbitrary state
$|\phi\rangle$ in the Fock space:
\begin{equation}
\langle \phi| Q_B|\psi\rangle+\langle \phi|\psi *\psi\rangle=0,
\end{equation}
and with the solution itself:
\begin{equation}
\langle \psi| Q_B|\psi\rangle+\langle \psi|\psi *\psi\rangle=0.
\label{EOM-SELF}
\end{equation}
Furthermore, the energy density calculated from the solution is
expected to cancel D25-brane tension \cite{9911116},
\begin{equation}
{\cal E}[\psi]=\frac{1}{g^2}\left[\frac{1}{2} \langle\psi|Q_B
|\psi \rangle +\frac{1}{3}\langle \psi|\psi *\psi\rangle \right],
\hspace{5mm}\frac{\cal E}{T_{25}}=-1,
\label{tension}
\end{equation}
with $T_{25}=\frac{1}{2\pi^2 g^2}$, where we have assumed, without
losing generality, that the brane has unit volume.

\begin{figure} \centerline{\hbox{\epsfig{figure=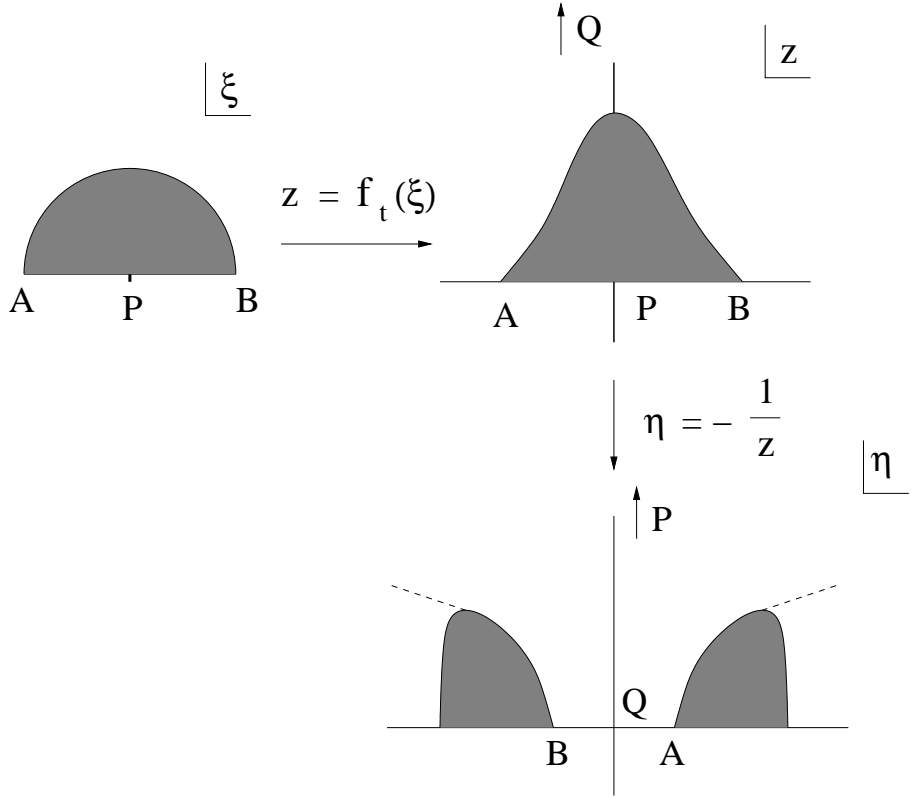,
width=12cm}}} \caption{The canonical half disk on the $\xi$ plane
is the local coordinate patch  with $\xi(P)=0$ being the puncture.
This half disk is mapped to some region around the origin on the
upper half $z$ plane, with the middle point of the boundary $Q$
lying at infinity. On the $\eta=I(z)=-\frac{1}{z}$ representation,
$Q$ lies at the origin and $P$ lies at infinity.} \label{rep}
\end{figure}

A general twist invariant star algebra projector $|P\rangle$ can
be defined on the upper half $z$ plane representation by
\footnote{For reviews of star algebra projectors and their various
representations, one can refer to \cite{0202151},
\cite{0006240}-\cite{0207001}.}:
\begin{equation}
\langle\phi|P\rangle=\langle f\circ \phi(0)\rangle_{\mathbb{H}_z}.
\end{equation}
$f(\xi)$ is fixed up to a scale by $f(0)=0$ and the twist
invariant condition $f(1)=-f(-1)$. A regulated projector
$|P_t\rangle$ is then defined by:
\begin{equation}
\langle\phi|P_t\rangle=\langle f_t\circ
\phi(0)\rangle_{\mathbb{H}_z},
\end{equation}
with the puncture located at $z=0$ and the middle point of the
boundary lying at $z=\infty$. $f_t(1)=-f_t(-1)$ fixes $f_t(\xi)$
up to a scale. We can set this scale ambiguity as the same as that
of $f(\xi)$. Therefore, we have
\begin{equation}
\lim_{t\to 0}f_t(\xi)=f(\xi), \hspace{5mm} \lim_{t\to 0}
\langle\phi|P_t\rangle= \langle\phi|P\rangle= \langle I\circ
f\circ \phi(0)\rangle_{\mathbb{H}_z}.
\end{equation}
The map $f_t(\xi)$, or equivalently, the shape of the local
coordinate patch on the upper half $z$ plane carries all the
information of the regulated projector. Since we will insert the
operators on the middle point of the boundary, the following
$\eta=I(z)=-\frac{1}{z}$ representation is useful:
\begin{equation}
\langle\phi|P_t\rangle=\langle I\circ f_t\circ
\phi(0)\rangle_{\mathbb{H}},
\end{equation}
which maps the middle point of the boundary to $\eta=0$, as
illustrated in figure (\ref{rep}). The star product of two
projectors can be easily done in $\hat z$ representation. In this
representation, the image of the local coordinate patch is the
full strip $\Re(\hat z)\leq \pi/4$, $\Im (\hat z)\geq 0$ and the
middle point of the string is mapped to $\hat z=i\infty$ under the
map $\hat z=\arctan\xi$. One removes the local coordinate patch of
the first projector, and glues the right half string of the first
projector with the left half string of the second projector. Then
we can map the glued surface in $\hat z$ representation to $\eta$
representation and obtain:
\begin{equation}
\langle\phi|P_t*P_t\rangle=\langle I\circ \tilde f_t\circ\phi
(0)\rangle_{\mathbb{H}}.
\end{equation}
In principle, $\tilde f_t(\xi)$ can be derived from $f_t(\xi)$,
but this is a nontrivial task. So far, we only know $\tilde
f_t(\xi)$ for sliver and butterfly states \cite{0310264, 0202139}.
The ambiguity of $\tilde f_t(\xi)$ is fixed up to a scale by
$\tilde f_t(0)=0$ and $\tilde f_t(1)=-\tilde f_t(-1)$. In the
$\eta$ representation, we denote:
\begin{equation}
\eta=I\circ f_t (\xi),\hspace{5mm} \eta'=I\circ\tilde f_t (\xi).
\end{equation}
From the construction of the star product of projectors mentioned
above, there exists a map:
\begin{equation}
\eta'=g_t(\eta) \label{trans}
\end{equation}
which tells us where the inserted operators for each single
projector are mapped on the glued $\eta'$ plane. This map is a one
to two map. $g_t(0)$ has two values with opposite signs which
represent the locations of the two insertions on the glued $\eta'$
surface. We will choose $g_t(0)>0$ without losing generality.

\subsection{The leading order solution}
At the first order, the $c$ ghost is inserted at the boundary
midpoint of the regulated projector. We use the notation
$|P_t(c)\rangle$ to denote this operator insertion. Therefore, on
the $\eta$ representation,
\begin{equation}
\langle \phi|P_t(c)\rangle=\langle I\circ f_t\circ \phi(\xi=0)\,
c(\eta=0) \rangle_{\mathbb{H}},
\end{equation}
for an arbitrary state $|\phi\rangle$ in the Fock space. Since the
BRST transformation of $c$ ghost is simply $Q_B\circ c=c\partial
c$, the kinetic term in the action (\ref{action}) gives:
\begin{eqnarray}
\label{GeneralKinetic} \langle\phi|Q_B|P_t(c)\rangle&=&
\langle I\circ f_t\circ \phi(0)\, c\p c(0)\rangle_{\mathbb{H}}\nonumber \\
&=&\langle I\circ f\circ \phi(0)\, c\p
c(0)\rangle_{\mathbb{H}}+{\cal O}(t),
\end{eqnarray}
where we have used the fact that one has the freedom to choose the
regulation parameter $t$ to make $f_t(\xi)=f(\xi)+{\cal O}(t)$.
For the cubic term, one has to glue two projectors, each of which
has $c$ ghost inserted at $\eta=0$. Therefore, one has to
transform $c(0)$ in each $\eta$ surface onto the glued $\eta'$
surface,
\begin{equation}
\langle\phi|P_t(c)*P_t(c)\rangle=\frac{1}{g_t'(0)^2} \langle
I\circ \tilde f_t\circ \phi(0) c(-g_t(0))
c(g_t(0))\rangle_{\mathbb{H}},
\end{equation}
with $g_t(0)>0$. The order of the two operators $c(-g_t(0))$ and
$c(g_t(0))$ is arranged to keep the orientation of the surface
unchanged after mapping\footnote{If we had choose $g_t(0)<0$, the
order of the two $c$ ghost insertions should be reversed.}. For
any projector, we have $\tilde f_t(\xi)=k f(\xi)+{\cal
O}(t^\alpha)$ with $\alpha>0$ and $k$ being the scale constant. If
$\tilde f_t(\xi)$ is analytic with respect to $t$, $\alpha$ is
some positive integer. Therefore, $g_t(0)\sim {\cal O}(t^\beta)$
with $\beta>0$. Thus, both $c(-g_t(0))$ and $c(g_t(0))$ approach
$c(0)$ as $t\to 0$. We need to evaluate the OPE of the two $c$
insertions.
\begin{equation}
\label{GeneralCubic} \langle\phi|P_t(c)*P_t(c)\rangle=\frac{2
g_t(0)}{k\,g_t'(0)^2}\Big(\langle I\circ f\circ \phi(0)\, c\p
c(0)\rangle_{\mathbb{H}}\Big) +{\cal O}(t^{\alpha+\beta})+{\cal
O}(t^{3\beta}).
\end{equation}
Suppose on the $\eta$ representation, the leading order solution
takes the form:
\begin{equation}
|\psi^{(0)}\rangle=-a_t|P_t(c)\rangle,
\end{equation}
where $a_t$ is some constant. From the equation of motion
$\langle\phi|Q_B|\psi\rangle+\langle\phi|\psi*\psi\rangle=0$,
comparing equations (\ref{GeneralKinetic}) and
(\ref{GeneralCubic}), one can identify $a_t={k g_t'(0)^2}/{2
g_t(0)}$. It is straightforward to see that $a_t$ is scale
invariant. Therefore, we simply choose $k=1$ in our paper and
obtain:
\begin{equation}
a_t=\frac{g_t'(0)^2}{2g_t(0)},
\end{equation}
provided\footnote{This condition is satisfied by sliver state and
butterfly state, where $\alpha=1$ and $\beta=\frac{1}{2}$. It is
not clear what constraints it imposes on projectors.} $\beta<1$
and $\beta<\alpha$ because the equation of motion based on this
solution now reads:
\begin{equation}
\langle\phi|Q_B|\psi^{(0)} \rangle+\langle\phi|\psi^{(0)}
*\psi^{(0)}\rangle={\cal O}(t^{1-\beta})+{\cal
O}(t^{\alpha-\beta})+{\cal O}(t^\beta).
\end{equation}
Since $g_t'(0)$ must be real from the orientations of the surface
before and after mapping, $a_t$ is positive. Therefore, the
leading order solution in $\eta$ representation is:
\begin{equation}
\label{General1st}
|\psi^{(0)}\rangle=
-\frac{g_t'(0)^2}{2g_t(0)}|P_t(c)\rangle.
\end{equation}
We see that as $t\to 0$, the coefficient
$-\frac{g_t'(0)^2}{2g_t(0)}$ becomes singular for any projector.

\subsection{The energy density for the leading order
solution}

\begin{figure}
\centerline{\hbox{\epsfig{figure=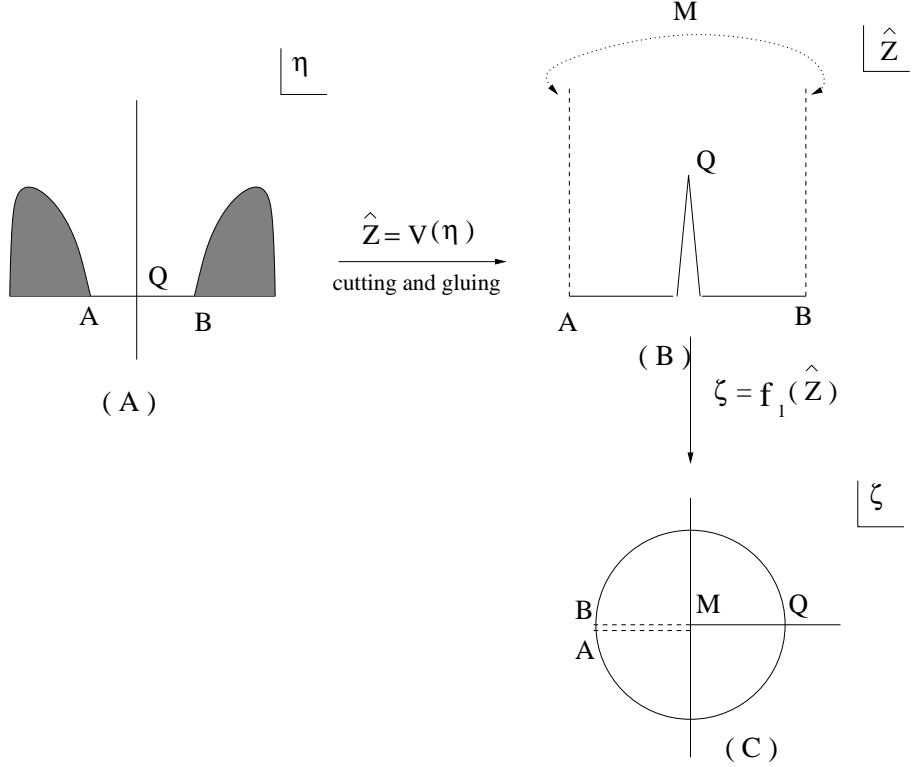, width=12cm}}}
\caption{Figure (A) is a general twist invariant regulated
projector in the $\eta$ representation. This projector is mapped
to $\hat z$ representation by $\hat z=v(\eta)=\arctan\circ(I\circ
f_t)^{-1}(\eta)$. After cutting the local coordinate patch and
identify the left half string $AM$ with the right half string
$BM$, figure (B) is obtained. Then figure (B) is mapped to a unit
circle in $\zeta$ plane by $f_1(\hat z)$ in figure (C). $M$ is the
string middle point, which lies at $i\infty$ in $\hat z$
representation and mapped to the origin in $\zeta$ plane. $Q$ is
the boundary middle point, which is mapped to $\zeta=1$ on the
unit disk.} \label{single}
\end{figure}

In the last subsection, we obtained the leading order solution for
general star algebra projectors. We want to use this solution to
calculate the energy density. In the calculations,  one encounters
the inner products of regulated projectors. Therefore, we first
map the regulated projector in $\eta$ representation to $\hat z$
representation. After gluing the projectors in $\hat z$
representation, one can map the resulted surface to a unit disk
and then evaluate the expectation values\footnote{We adopt the
normalization $\langle c_{-1}c_0c_1\rangle_D=1$ in this paper.}.

Let's first consider a single regulated projector on $\eta$
representation. The map from $\eta$ representation to $\hat z$
representation is:
\begin{equation}
\hat z=\arctan\circ(I\circ f_t)^{-1}(\eta)\equiv v(\eta).
\label{v}
\end{equation}
After cutting the local coordinate patch and identifying the right
half string with the left half string, we next map this surface to
a unit disk on the $\zeta$ plane with the middle point of boundary
$Q$ mapped to $\zeta=1$ and the middle point of the string $M$
mapped to $\zeta=0$ by:
\begin{equation}
\zeta=f_1(\hat z)=f_1\circ v(\eta)\equiv h(\eta),
\label{Z}
\end{equation}
as illustrated in figure (\ref{single}). Since AM is glued with
BM, $f_1(\hat z)$ is periodic with respect to $\hat z$. Usually,
it is not very hard to figure out $f_1(\hat z)$ for a particular
projector. For an instance, for the generalized butterfly states
\cite{0202151} parametrized by $0\leq\alpha\leq 2$, after cutting
and gluing, one can translate and rescale the surface via
$\frac{2\alpha}{2-\alpha}(\hat z-\pi/4)$ to obtain the same
surface as a regulated butterfly state. The map for a regulated
butterfly state from the $\hat z$ representation to the upper half
plane is $f_t(\hat z)=\frac{\tan\hat z}{\sqrt{1+t^2\tan^2\hat z}}$
with $t$ being the regulation parameter. Therefore, for the
generalized butterfly states, we find:
\begin{equation}
f_1(\hat z)=\frac{\sqrt{1+t'^2
\tan^2(\frac{2\alpha}{2-\alpha}(\hat
z-\pi/4))}+i\sqrt{1-t'^2}\tan(\frac{2\alpha}{2-\alpha}(\hat
z-\pi/4))}{-\sqrt{1+t'^2 \tan^2(\frac{2\alpha}{2-\alpha}(\hat
z-\pi/4))}+i\sqrt{1-t'^2}\tan(\frac{2\alpha}{2-\alpha}(\hat
z-\pi/4))}, \label{general f1}
\end{equation}
with $t'=\tanh(\frac{2\alpha}{2-\alpha}\tanh^{-1} t)$, where $t$
is the regulation parameter. We will simply assume this map is
given for any twist invariant projector.

\begin{figure}
\centerline{\hbox{\epsfig{figure=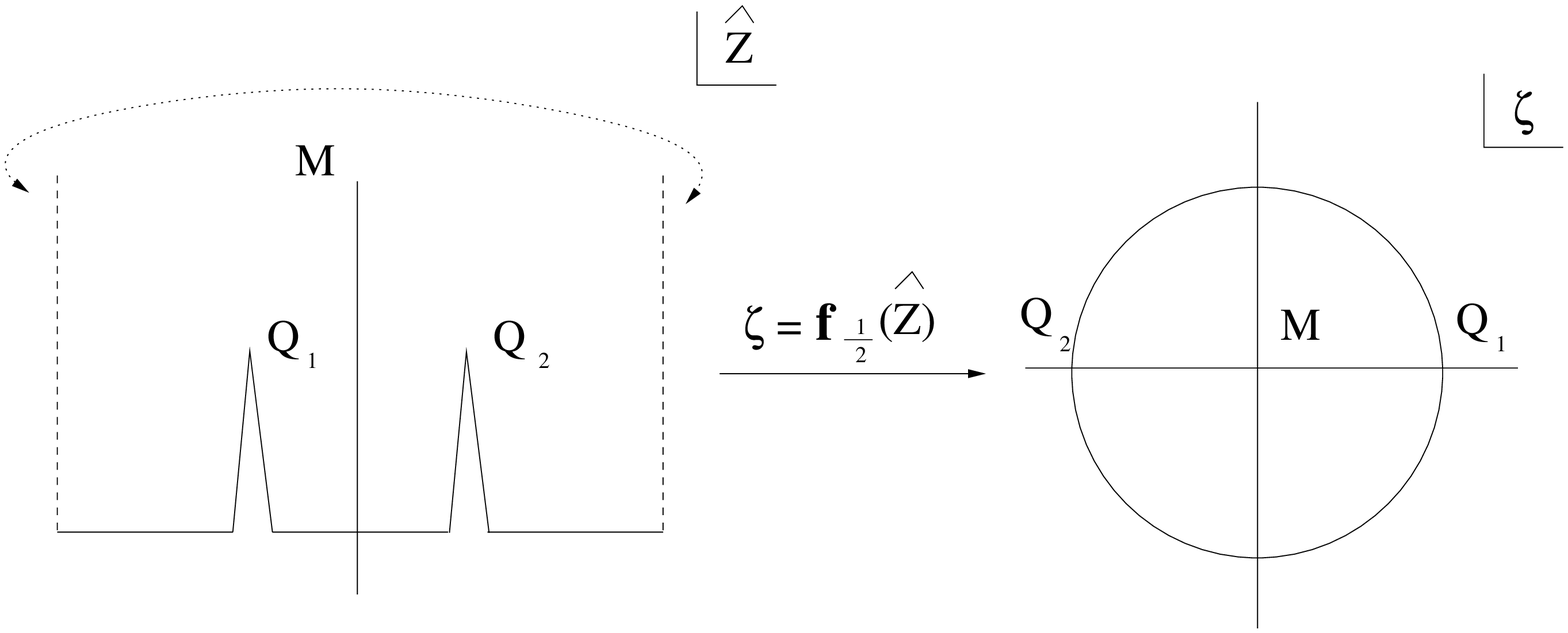, height=6cm}}}
\caption{Mapping the inner product of two projectors in $\hat z$
representation to a unit disk in $\zeta$ plane.} \label{two}
\end{figure}

Then the map\footnote{This idea was first discussed in
\cite{0310264} for regulated butterfly states.} from the inner
product of two regulated projectors in the $\hat z$ representation
to a unit disk in the $\zeta$ plane is $f_{1/2}(\hat
z)=\sqrt{f_1(\hat z)}$, as illustrated in figure (\ref{two}) .
From the periodicity of $f_1(\hat z)$, we can see that
$f_1(Q_1)=f_1(Q_2)$ as well as $f_{1/2}(Q_1)=-f_{1/2}(Q_2)=1$,
where $Q_1$ and $Q_2$ are the boundary middle points of the two
regulated projectors. For the inner product of three regulated
projectors, the map is $f_{1/3}(\hat z)=f_1^{1/3}(\hat z)$ with
$f_1(Q_1)=f_1(Q_2)=f_1(Q_3)$ as well as $f_{1/3}(Q_1)=1$,
$f_{1/3}(Q_2)=\exp(i 2\pi/3)$, $f_{1/3}(Q_3)=\exp(i 4\pi/3)$, with
$Q_1$, $Q_2$ and $Q_3$ being the boundary middle points of the
three regulated projectors. Therefore,
\begin{eqnarray}
\langle P_t(c)|Q_B|P_t(c)\rangle &=& \frac{A^2}{f_{1/2}'(\hat
z(Q_1))f_{1/2}'(\hat z(Q_2))} \langle c(1) c\p
c(-1)\rangle_D\nonumber\\
&=&\frac{-4A^2}{f_{1/2}'(\hat z(Q_1))f_{1/2}'(\hat z(Q_2))},
\end{eqnarray}
with
\begin{equation}
A=-\frac{1}{v'(\eta=0)}\frac
{g_t'(0)^2}{2g_t(0)}=-\frac{1}{[\arctan\circ (I\circ
f_t)^{-1}(\eta=0)]'}\frac {g_t'(0)^2}{2g_t(0)}, \label{A}
\end{equation}
from equation (\ref{Z}). The cubic interaction is:
\begin{equation}
\langle P_t(c)|P_t(c)*P_t(c)\rangle=\frac{i
3\sqrt{3}A^3}{f_{1/3}'(Q_1)f_{1/3}'(Q_2)f_{1/3}'(Q_3) }.
\end{equation}
From the periodicity of $f_1(\hat z)$, we have
\begin{equation}
f_{1/2}'(Q_1)f_{1/2}'(Q_2)=\frac{f_1'^2(Q_1)}{4f_{1/2}(Q_1)f_{1/2}(Q_2)}
=-\frac{f_1'^2(Q_1)}{4},
\end{equation}
for the kinetic term and
\begin{equation}
f_{1/3}'(Q_1)f_{1/3}'(Q_2)f_{1/3}'(Q_3)=\frac{f_1'^3(Q_1)}{27},
\end{equation}
for the cubic interaction. It sometimes happens that
$\frac{1}{[v(\eta=0)]'}$, which appears in the expression of $A$
and $f_1'(Q_1)$, is singular for some projectors. Actually, this
occurs for the butterflies. This is because $\psi^{(0)}$ is not
well defined in $\hat z$ representation\footnote{Thanks to Okawa
for pointing out this fact.} for some regulated projectors.
However, the map $h(\eta)$ to a unit disk defined in (\ref{Z}) is
well defined for any projector. Therefore, we write:
\begin{equation}
\label{FinalKinetic} \lim_{t\to 0}\langle
P_t(c)|Q_B|P_t(c)\rangle=\lim_{t\to 0}\frac{16}{h'(\eta=0)^2}
\Big(\frac {g_t'(0)^2}{2g_t(0)}\Big)^2,
\end{equation}
\begin{equation}
\label{FinalCubic} \lim_{t\to 0}\langle
P_t(c)|P_t(c)*P_t(c)\rangle=-\lim_{t\to 0}\frac{i
3^{9/2}}{h'(\eta=0)^3}\Big(\frac {g_t'(0)^2}{2g_t(0)}\Big)^3,
\end{equation}
where we have set $t\to 0$ and only kept the leading order. Let us
define:
\begin{equation}
b_t\equiv -ih'(\eta=0).
\end{equation}
One can justify $b_t$ is positive because $h'(0)=i\times$ (a real
number) by considering the orientations of the surface before and
after mapping. Thus, the ratio between the kinetic term and the
cubic interaction is:
\begin{equation}
\label{ratio} R=\lim_{t\to 0}\frac{\langle
\psi^{(0)}|Q_B|\psi^{(0)}\rangle} {\langle
\psi^{(0)}|\psi^{(0)}*\psi^{(0)}\rangle}=-2^4\cdot
3^{-9/2}\lim_{t\to 0}\frac{b_t}{a_t},
\end{equation}
which is always negative. The ratio of the energy density at this
solution to D25-brane tension is:
\begin{equation}
\label{Energy}
\frac{\cal E}{T_{25}}=2\pi^2\lim_{t\to
0}\Big[-8\Big(\frac {a_t}{b_t}\Big)^2+ 3^{7/2}\Big(\frac
{a_t}{b_t}\Big)^3 \Big].
\end{equation}
One can see that both quantities only depend on one positive
number: $\lim_{t\to 0}\frac{b_t}{a_t}$. A universal relationship
between $R$ and ${\cal E}/T_{25}$ follows:
\begin{equation}
\frac{\cal E}{T_{25}}=- \frac{\pi^2}{3}\left(\frac{4}{3\sqrt
3}\right)^6\frac{3R+2}{R^3}\simeq -0.684616\times
\frac{3R+2}{R^3},
\end{equation}
independent of the details of the projectors. The number $-0.6846$
is the famous one obtained in the first order level truncation
calculations. The quantity $\frac{3R+2}{R^3}$ acquires its maximum
at the ideal value $R=-1$. Therefore, we can conclude that for any
twist invariant projector, at the leading order, the energy
density can account for at most $68.4616\%$ of the D25-brane
tension.

\sectiono{The leading order solution for regulated sliver and
butterfly states.}

In this section, we will apply the results obtained in last
section to regulated sliver and butterfly states. The results for
the regulated butterfly state are exactly the same as those
obtained in \cite{0311115}. We will verify the solution for the
regulated sliver state by checking the equation of motion
(\ref{EOM-SELF}) and the ratio of energy density to D25-brane
tension in (\ref{tension}), when contracted with itself.

\subsection{Results for the regulated slivers}

\begin{figure}
\centerline{\hbox{\epsfig{figure=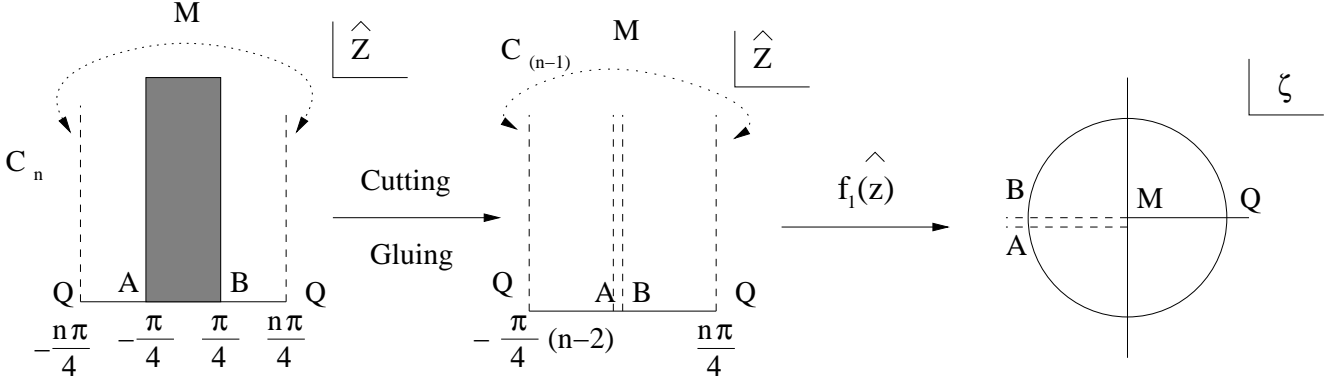, height=5cm, }}}
\caption{A wedge state $|n\rangle$ is defined by the surface $C_n$
on $\hat z$ representation. After cutting the local coordinate
patch and gluing left half string with right half string, one
obtains $C_{n-1}$ surface on $\hat z$. $f_1(\hat z)$ maps
$C_{n-1}$ to a unit disk with that the image of $Q$ is $\zeta=1$
and the image of $M$ is $\zeta=0$. } \label{sliver}
\end{figure}

The regulated sliver state is a wedge state parametrized by $n$.
As $n\to \infty$, one obtains the real sliver state $|\Xi\rangle$.
For a single wedge state $|n\rangle$,
\begin{equation*}
\langle \phi|n\rangle=\langle  f\circ \phi(0)\rangle_{C_n}=\langle
I\circ f_n\circ f\circ \phi(0)\rangle_{\mathbb{H}},
\end{equation*}
with $f(\xi)=\arctan\xi$  and $f_n(\hat
z)=\frac{n}{2}\tan\left(\frac{2}{n} \hat z\right)$ \cite{0008252},
where we have already mapped the middle point of the boundary to
$\eta=0$  as required in equation (\ref{trans}). The surface $C_n$
is a cylinder with perimeter $\frac{n\pi}{2}$ on upper half plane,
defining the wedge state $|n\rangle$ in $\hat z$ representation,
as illustrated in figure (\ref{sliver}).
For the star product of two wedge states:
\begin{equation*}
\langle \phi|n*n\rangle=\langle  f\circ
\phi(0)\rangle_{C_{2n-1}}=\langle  I\circ \tilde f_{n}\circ f\circ
\phi(0)\rangle_{\mathbb{H}},
\end{equation*}
where $\tilde f_{n}(\hat z)= f_{2n-1}(\hat
z)=\frac{2n-1}{2}\tan\left(\frac{2}{2n-1} \hat z\right)$. The map
from a single wedge state to the glued one in $\eta$
representation is:
\begin{equation*}
g_n(\eta)=\frac{2}{(2n-1)\tan\Big[\frac{n}{2n-1}\arctan\frac{2}{n\eta}\big]}.
\end{equation*}
Thus,
\begin{equation*}
a_n=\frac{g_n'(0)^2}{2g_n(0)}=\frac{n^4}{4(2n-1)^3\cos\frac{n\pi}{4n-2}
\sin^3\frac{n\pi}{4n-2}}=\frac{n}{8}+{\cal O}(1).
\end{equation*}
From equation (\ref{General1st}), in $\eta$ representation, the
leading order solution is:
\begin{equation}
|\psi^{(0)}\rangle=-\frac{n}{8}|n(c)\rangle.
\end{equation}
It is straightforward to calculate $v(\eta)$:
\begin{equation}
\hat z=v(\eta)=\arctan\circ(I\circ f_n\circ
f)^{-1}(\eta)=-\frac{n}{2}\arctan\frac{2}{n\eta}.
\label{sliverv}
\end{equation}
For a wedge state $|n\rangle$ in the $\hat z$ representation with
boundary midpoint at $\hat z(Q)=\frac{n\pi}{4}$ and string
midpoint at $\hat z(M)=i\infty$, after cutting the local
coordinate patch and gluing the left half string with the right
half string, one obtains the surface $C_{n-1}$. Then, as
introduced in equation (\ref{Z}),
\begin{equation}
f_1(\hat z)=\exp\left(i\frac{4}{n-1}(\hat z-n\pi/4)\right),
\label{sliverf1}
\end{equation}
maps this $C_{n-1}$ surface to a unit disk with the image of $\hat
z(Q)$ being $\zeta=1$ and the image of $\hat z(M)$ being the
origin $\zeta=0$, as illustrated in figure (\ref{sliver}). This
map can also be obtained readily from equation (\ref{general f1})
by setting $\alpha=2/n$ as well as $t=0$. Therefore,
\begin{equation}
b_n=-ih'(0)=-i(f_1\circ v(\eta=0))'=\frac{n^2}{n-1}.
\end{equation}
From equations (\ref{FinalKinetic}) and (\ref{FinalCubic}), the
leading order solution gives:
\begin{equation}
\lim_{n\to\infty} \langle \psi^{(0)}|Q_B|\psi^{(0)}\rangle =
-\frac{1}{4},
\end{equation}
\begin{equation}
\lim_{n\to\infty} \langle \psi^{(0)}|\psi^{(0)}*\psi^{(0)}\rangle=
\left(\frac{3\sqrt 3}{8}\right)^3.
\end{equation}
The ratio of the kinetic term to the cubic interaction which is
expected to be $-1$ from the equation of motion reads:
\begin{equation}
R=\lim_{n\to\infty}\frac{\langle \psi^{(0)}|Q_B|\psi^{(0)}\rangle}
{\langle \psi^{(0)}|\psi^{(0)}*\psi^{(0)}
\rangle}=-2\left(\frac{4}{3\sqrt 3}\right)^3=-0.9124.
\end{equation}
The ratio of the energy density to D25-brane tension is:
\begin{equation}
\lim_{n\to\infty}\frac{{\cal
E}}{T_{25}}=2\pi^2\left(-\frac{1}{8}+\frac{3^{7/2}}
{8^3}\right)=-0.6645.
\end{equation}
Compared with the value $-0.6846$ obtained by level truncation at
the first tachyon level, this ratio is rather close.

\subsection{Solution of the regulated butterflies}

A regulated butterfly state $|B_t\rangle$ characterized by a
parameter $t\in [0,1)$ is defined by:
\begin{equation}
\langle \phi|B_t\rangle=\langle f_t(\xi)\circ
\phi(0)\rangle_{\mathbb{H}_z}, \hspace{5mm}
f_t(\xi)=\frac{\xi}{\sqrt{1+t^2\xi^2}},
\end{equation}
for an arbitrary state $|\phi\rangle$ in the Fock space. As $t\to
1$, one obtains the exact butterfly state $|B\rangle$. The
solution based on regulated butterfly states was calculated in
\cite{0311115} by Okawa to the next to leading order. From
equation (2.33)\footnote{The notations in \cite{0311115} are
different from ours. There, the author defined $z'=I\circ
f_t(\xi)$ and $z=\tilde f_t(\xi)$.} in \cite{0311115},
$g_t(0)=\beta$ and $g_t'(0)=\sqrt\frac{48}{(1-t^4)(1-a^2)(3+a^2)}
\beta^2$ with $\beta=\frac{2}{9-a^2}\sqrt{(1-a^2)(3+a^2)}$ and
$a=\sqrt 3\tan\left(\frac{2}{3}\arctan t\right)$. Therefore,
\begin{equation}
a_t=\frac{g_t'(0)^2}{2g_t(0)}=\frac{3^{1/4}}{4\sqrt
2}\frac{1}{\sqrt{1-t}}+{\cal O}(\sqrt{1-t}).
\end{equation}
Thus, the leading order solution in $\eta$ representation is:
\begin{equation}
|\psi^{(0)}\rangle=-\frac{3^{1/4}}{4\sqrt 2}\frac{1}{\sqrt{1-t}}
|B_t(c)\rangle,
\end{equation}
exactly the same as equation (2.41) in \cite{0311115}. Given
$f_t(\xi)=\frac{\xi}{\sqrt{1+t^2\xi^2}}$, one obtains:
\begin{equation}
\hat z=v(\eta)=\arctan\circ (I\circ f_t)^{-1}(\eta)=\frac{\pi}{2}+
i\,\mbox{arctanh}\,\sqrt{t^2-\eta^2},
\end{equation}
with the normalization that the boundary middle point is mapped to
$\hat z=\frac{\pi}{2}+ i\,\mbox{arctanh}\,t$. The surface obtained
by cutting the local coordinate patch and gluing the left half
string with the right half string of a regulated butterfly state
in $\hat z$ representation is mapped to a unit disk on $\zeta$
plane by equation (\ref{general f1}) with $\alpha=1$:
\begin{equation}
f_1(\hat z)=\frac{\tan(2\hat z-\pi/2)-i\sqrt{1+q^2+q^2\tan^2(2\hat
z-\pi/2)}} {\tan(2\hat z-\pi/2)+i\sqrt{1+q^2+q^2\tan^2(2\hat
z-\pi/2)}},
\label{butterflyf1}
\end{equation}
with $q=\frac{2 t}{1-t^2}$. The string midpoint is mapped to
$\zeta(M)=0$ and the image of the boundary midpoint is located at
$\zeta(Q)=1$. Therefore,
\begin{equation}
b_t=-ih'(0)=-i(f_1\circ v(\eta=0)) '=4(1-t^4)^{-1/2}.
\end{equation}
From equations (\ref{FinalKinetic}), (\ref{FinalCubic}),
(\ref{ratio}) and (\ref{Energy}),
\begin{equation}
\lim_{t\to 1}\langle B_t(c)|Q_B|B_t(c)\rangle=-\frac{\sqrt 3}{8},
\end{equation}
\begin{equation}
\lim_{t\to 1}\langle B_t(c)|B_t(c)*B_t(c)\rangle=3^{21/4}
2^{-21/2},
\end{equation}
as well as
\begin{eqnarray}
R&=& -2^{15/2}\cdot 3^{-19/4},\nonumber \\
\frac{\cal E}{T_{25}}&=&2\pi^2\left(-\frac{\sqrt 3}{16}+ 3^{17/4}
2^{-21/2}\right)+{\cal O}(1-t).
\end{eqnarray}
All the results agree with those obtained in \cite{0311115}.

\sectiono{Subleading solution for a regulated sliver state}
In the last few sections, analytical expressions for the leading
order solution were given for general star algebra projectors. In
this section, we will calculate the next to leading order solution
based on a regulated sliver state in the $\hat z$ representation.
The reason for this choice is that projector gluing can be easily
done in this representation and the state is well defined in the
$\hat z$ representation for wedge states. One can transform the
solution to other representations by conformal maps.

From last section, the leading order solution based on wedge state
is $|\psi^{(0)}\rangle=-\frac{n}{8}|n(c)\rangle$. With the
conformal map (\ref{sliverv}), one can verify that this solution
is $|\psi^{(0)}\rangle=-\frac{1}{2n}|n(c)\rangle_{\hat z}$ in the
$\hat z$ representation, where we use the subscript $\hat z$ to
denote that the operator is inserted at $\hat z=n\pi/4$, the
boundary midpoint of the wedge state $|n\rangle$ on $\hat z$
representation. Two contributions should be considered in the next
order. First, $f_n(\hat z)$ and $\tilde f_n (\hat z)$ have
different expansions in the next to leading order. One should also
take into account higher orders in the OPE of the two $c$ operator
insertions. Almost parallel to the regulated butterfly situation,
the next to leading order solution takes the form:
\begin{equation}
|\psi^{(2)}\rangle=\frac{x}{n}|n(c)\rangle_{\hat
z}+n\Big( u|n(\p^2c)\rangle_{\hat z}+v|n(cT^m)\rangle_{\hat z}
+w|n(:bc\p c:)\rangle_{\hat z}\Big), \label{2nd order}
\end{equation}
with all the operators inserted at $\hat z=\frac{n\pi}{4}$, the
middle point of the boundary. One should note that we are working
in the $\hat z$ representation. That's why the leading order has a
coefficient of $\frac{1}{n}$, whereas next to leading order has a
coefficient of $n$. If we map the solution to the upper half
$\eta$ plane, the coefficient of the leading order will be $n$ and
coefficient of the next order will be $\frac{1}{n}$. The BRST
transformations of these operators are:
\begin{eqnarray}
\label {BRST} Q_B\circ c= c\p c,&\hspace{5mm}&
Q_B\circ \p^2 c=\p c\p^2c+c\p^3c,\nonumber\\
Q_B\circ cT^m=-\frac{13}{6}c\p^3 c-c\p c T^m,&\hspace{5mm}&
Q_B\circ :bc\p c:=\frac{2}{3}c\p^3 c-\frac{3}{2}\p c\p^2 c +c\p c
T^m.
\end{eqnarray}
The conformal transformation rules of some relevant non-tensor
operators are summarized in Appendix A.

\subsection{Calculation of the kinetic term}

We first calculate the kinetic term. From eqn. (\ref{BRST}),
\begin{eqnarray}
\langle \phi|Q_B|\psi^{(2)}\rangle&=&\langle f\circ
\phi(0)\, Q_B\circ\psi^{(2)}\left(\frac{n\pi}{4}\right)\rangle_{C_n}
\nonumber\\
&=&\frac{x}{n} \langle f\circ \phi\, c\p c \left(
\frac{n\pi}{4} \right)\rangle_{C_n} + n\bigg \{(u-\frac{3}{2}w)
\langle f\circ\phi\, \p c\p^2 c\left( \frac{n\pi}{4}
\right) \rangle_{C_n} \nonumber\\
&&+ (u-\frac{13}{6}v+\frac{2}{3} w) \langle f\circ \phi\, c\p^3
c\left( \frac{n\pi}{4} \right)\rangle_{C_n} \nonumber\\
&&+(w-v)\langle f\circ\phi\, c\p c T^m \left( \frac{n\pi}{4}
\right)\rangle_{C_n}\bigg\},
\end{eqnarray}
where $C_n$ represents a wedge state $|n\rangle$ in the $\hat z$
representation, as defined in last section. In order to determine
the coefficients $x$, $u$, $v$ and $w$, one has to compare this
result with that from cubic interaction by the virtue of equation
motion. A convenient choice is to evaluate all the expectation
values on upper half plane. From equation (\ref{TRANS}), under
$h_n(\hat z)\equiv I\circ f_n(\hat z)=-\frac{2}{n\tan(2\hat z/n)}$
which maps $C_n$ to upper half $\eta$ plane,
\begin{eqnarray}
\langle \phi|Q_B|\psi^{(2)}\rangle &=&
n\left(\frac{x}{4}-2u+\frac{13}{3}v-\frac{4}{3}w\right)
\langle h_n\circ f\circ \phi(0)\, c\p c(0)\rangle_{\mathbb{H}} \nonumber \\
&&+\frac{4}{n}\left(u-\frac{3}{2}w\right) \langle h_n\circ f\circ
\phi(0)\, \p c\p^2 c(0)\rangle_{\mathbb{H}}\nonumber\\
&&+\frac{4}{n}\left(u-\frac{13}{6}v+\frac{2}{3}w\right) \langle
h_n\circ f\circ \phi(0)\, c\p^3 c(0)\rangle_{\mathbb{H}} \nonumber\\
&&+\frac{4}{n}(w-v)\langle h_n\circ f\circ \phi(0)\, c\p c
T^m(0)\rangle_{\mathbb{H}}. \label{uph}
\end{eqnarray}
On the other hand, since
\begin{equation}
f_n(\hat z)=\frac{n}{2}\tan\left(\frac {2\hat z}{n}\right) = \hat
z+\frac{1}{3} \left(\frac{2}{n}\right)^2 \hat z^2+{\cal O}\left(
\frac{1}{n^4}\right), \label{f}
\end{equation}
we have,
\begin{eqnarray}
\label{fnEXP} \langle h_n\circ f\circ \phi(0) c\p
c(0)\rangle_{\mathbb{H}}&=& \langle I\circ
f_n\circ f\circ\phi(0)c\p c(0)\rangle_{\mathbb{H}}\nonumber\\
&=&\langle I\circ f\circ\phi(0)c\p
c(0)\rangle_{\mathbb{H}}\nonumber\\
&&+\frac{4}{3n^2}\bigg\{\frac{2}{3} \langle I\circ
f\circ\phi(0)c\p^3 c(0)\rangle_{\mathbb{H}} -\frac{3}{2} \langle
I\circ f\circ\phi(0)\p c\p^2
c(0)\rangle_{\mathbb{H}}\nonumber\\
&&+\langle I\circ f\circ\phi(0)c\p
cT^m(0)\rangle_{\mathbb{H}}\bigg\}+{\cal
O}\left(\frac{1}{n^4}\right).
\end{eqnarray}
Therefore, up to the order of ${\cal O}\left(\frac{1}{n}\right)$,
our final expression for the kinetic term is:
\begin{eqnarray}
\label{2piont2} \langle \phi|Q_B|\psi^{(2)}\rangle&=&n\left(\frac{x}{4}-2u+\frac{13}{3}v-\frac{4}{3}
w\right)\langle \phi|\Xi (c\p c)\rangle\nonumber\\
&&+\frac{4}{n}\left(2u-\frac{x}{8}-\frac{13}{6}v-\frac{5}{6}
w\right)\langle \phi|\Xi (\p c\p^2 c)\rangle\nonumber\\
&&+\frac{4}{n}\left(\frac{x}{18}+\frac{5}{9}u-\frac{65}{54}v+\frac{10}{27}
w\right)\langle \phi|\Xi (c\p^3 c)\rangle\nonumber\\
&&+\frac{4}{n}\left(\frac{x}{12}-\frac{2}{3}u+\frac{4}{9}v+\frac{5}{9}
w\right)\langle \phi|\Xi (c\p cT^m)\rangle+{\cal O}\left(
\frac{1}{n^3}\right).
\end{eqnarray}

\subsection{Calculation of cubic interaction}
In the $\hat z$ representation, the surface corresponding to the
star product $|n*n\rangle$ of two wedge states is the surface
$C_{2n-1}$. The cubic term is:
\begin{equation}
\langle \phi|\psi^{(2)} *\psi^{(2)}\rangle=\langle f\circ\phi(0)\, \psi^{(2)}
(n\pi/4) \psi^{(2)} (-n\pi/4)\rangle_{C_{2n-1}}.
\end{equation}
Under
\begin{equation}
\eta=h_{2n-1}(\hat z)\equiv I\circ f_{2n-1}(\hat
z)=-\frac{2}{2n-1}\frac{1}{\tan(\frac{2\hat z}{2n-1})},
\end{equation}
which maps $C_{2n-1}$ to upper half $\eta$ plane, from equation
(\ref{TRANS}),
\begin{eqnarray*}
& &\frac{x}{n}c(\pm n\pi/4)+n\Big(u\p^2 c(\pm n\pi/4)+ vcT^m(\pm
n\pi/4)+w:bc\p c:(\pm n\pi/4)\Big)\\
&\to&nAc(\mp t)+B c(\mp t)\pm C\p c(\mp t)\\
&&+\frac{1}{n}\Big(D c(\mp t)\pm Gc(\mp t)+F\p^2 c(\mp t)+H
cT^m(\mp t)+I:bc\p c:(\mp t)\Big)+{\cal O}\left(
\frac{1}{n^2}\right),
\end{eqnarray*}
where
\begin{eqnarray}
A&=&\frac{3x+13v-13w}{6},\nonumber\\
B&=&\frac{-12x+\pi(-12u+13v+5w+3x)}{24},\nonumber\\
C&=&2u-3w,\nonumber\\
D&=&\frac{6x+\pi(-12u+13v+5w-3x)}{48},\nonumber\\
G&=&\frac{(2u-3w)(2-\pi)}{4},\hspace{5mm} F=2u,\hspace{5mm}
H=2v,\hspace{5mm} I=2w,
\end{eqnarray}
and
\begin{equation}
t\equiv h_{2n-1}\left(-\frac{n\pi}{4}\right) =
\frac{2}{2n-1}\frac{1}{\tan\left(\frac{n\pi}{4n-2}\right)},
\label{t}
\end{equation}
with $\pm t$ are the two boundary middle points of the two
regulated sliver state after gluing in $\eta$ representation.
Therefore, on upper half $\eta$ plane,
\begin{eqnarray*}
&&\langle \phi|\psi^{(2)}*\psi^{(2)}\rangle=\langle h_{2n-1}\circ f\circ \phi(0)
\bigg\{ nAc(- t)+B c(- t)+ C\p c(- t) \\
&&+\frac{1}{n}\Big(D c(- t)+ Gc(- t)+F\p^2 c(- t)+H cT^m(-
t)+I:bc\p c:(- t)\Big)\bigg\}\\
&&\times \bigg\{nAc(t)+B c(t)-C\p c(t)\\
&&+\frac{1}{n}\Big(D c(t)-Gc(t)+F\p^2 c(t)+H cT^m(t)+I:bc\p c:(
t)\Big)\bigg\}\rangle_{\mathbb{H}}.
\end{eqnarray*}
As $n\to\infty$, $t\to 0$ from equation (\ref{t}), we need to
calculate the OPE's of the operators. The relevant OPEs are
presented in Appendix B. Thus,
\begin{eqnarray*}
\langle \phi|\psi^{(2)}*\psi^{(2)}\rangle&=&n F_1\langle
h_{2n-1}\circ f\circ \phi\, c\p c(0)\rangle+\frac{1}{n}
F_2\langle h_{2n-1}\circ f\circ \phi\,\p c\p^2 c(0)\rangle\\
&+&\frac{1}{n}F_3 \langle h_{2n-1}\circ f\circ \phi\, c\p^3
c(0)\rangle +\frac{1}{n}F_4\langle h_{2n-1}\circ f\circ \phi\, c\p
cT^m(0)\rangle,
\end{eqnarray*}
with
\begin{eqnarray*}
F_1&=&2A^2-2AC-AI-\frac{1}{2}CI-\frac{1}{2}FI-\frac{3}{8}I^2+{\cal
O}\left(\frac{1}{n}\right)\\
F_2&=&-A^2+3AC-2AF-\frac{1}{2}AI-2C^2+2CF-\frac{1}{4}FI+\frac{15}{16}
I^2-\frac{CI}{4}+{\cal
O}\left(\frac{1}{n}\right)\\
F_3&=&\frac{1}{3}A^2-AC+2AF-\frac{1}{2}AI-\frac{1}{4}FI-\frac{7}{48}I^2-\frac{CI}{4}+{\cal
O}\left(\frac{1}{n}\right)\\
F_4&=&4HA-2CH-HI+{\cal O}\left(\frac{1}{n}\right).
\end{eqnarray*}
As in equation (\ref{fnEXP}), we should expand the term $\langle
h_{2n-1}\circ f\circ \phi c\p c(0)\rangle$, but with
$\frac{4}{n^2}$ replaced by $\frac{1}{n^2}$ since
$f_{2n-1}(z)=z+\frac{1}{3} \frac{1}{n^2} z^3+{\cal
O}\left(\frac{1}{n^3}\right)$ compared with equation (\ref{f}).
Finally, the cubic interaction is:
\begin{eqnarray}
\label{3piont2} \langle \phi|\psi^{(2)}*\psi^{(2)}\rangle&=&n
F_1\langle \phi|\Xi(c\p c)\rangle+\frac{1}{n}
(F_2-\frac{1}{2}F_1)\langle \phi|\Xi(\p c\p^2
c)\rangle+\frac{1}{n}(F_3+\frac{2}{9}F_1) \langle \phi|\Xi(c\p^3
c)\rangle\nonumber\\
&& +\frac{1}{n}(F_4+\frac{1}{3}F_1)\langle \phi|\Xi(c\p
cT^m)\rangle+{\cal O}\left(\frac{1}{n^2}\right).
\end{eqnarray}
From the equation of motion, with equations (\ref{2piont2}) and
(\ref{3piont2}), we need to solve the following equations for
$x,u,v$ and $w$:
\begin{eqnarray}
-F_1&=&\frac{x}{4}-2u+\frac{13}{3}v-\frac{4}{3}
w\nonumber\\
-(F_2-F_1/2)&=&4\left(2u-\frac{x}{8}-\frac{13}{6}v-\frac{5}{6}
w\right)\nonumber\\
-(F_3+\frac{2}{9}F_1)&=&4\left(\frac{x}{18}+\frac{5}{9}u-\frac{65}
{54}v+\frac{10}{27}w\right)\nonumber\\
-(F_4+\frac{1}{3}F_1)&=&4\left(\frac{x}{12}-\frac{2}{3}u+\frac{4}{9}v+\frac{5}{9}
w\right).
\end{eqnarray}
There are four real valued nontrivial solutions obtained by
MATHEMATICA, presented in Table (\ref{solutions}).
\begin{table}[ht]
\caption{The next-to-leading order solutions based on sliver
state.} \label{solutions}
\begin{center}
{\renewcommand\arraystretch{1.1}
\begin{tabular}{|c|c|c|c|}
  \hline
  $x$ & $u$ & $v$ & $w$ \\
  \hline
  $2.62993$ & $-2.60731$ & $0.703291$ & $9.42887$ \\
  \hline
  $-0.493996$ & $0.00603476$ & $0.00564685$ & $0.0553046$ \\
  \hline
  $0.12831$ & $0.213274$ & $0.147305$ & $0.112232$ \\
  \hline
  $0$ & $6.90548\times 10^{-13}$ & $4.60365\times 10^{-13}$ & $4.60365\times
  10^{-13}$\\
  \hline
\end{tabular}}
\end{center}
\end{table}

The value of $x$ of the second solution is quite close to the
leading order result $x=-0.5$.

\subsection{Verifying the results}
We want to calculate the quantities $\langle \psi^{(2)}
|\psi^{(2)}\rangle$ and $\langle \psi^{(2)}
|\psi^{(2)}* \psi^{(2)}\rangle$ just as what we
did in the leading order solution.

The surface of the inner product of two wedge states on $\hat z$
representation is $C_{2n-2}$. The two punctures where operators
inserted are located at $\hat z=0$ and $\hat z=\frac{n-1}{2} \pi$.
\begin{equation}
g(\hat z)=\exp\left(i\frac{2\hat z}{n-1}\right),
\end{equation}
maps $C_{2n-2}$ to a unit disk. Under this map, the operators
inserted at $\hat z=0$ are transformed to:
\begin{eqnarray}
&&\frac{x}{n}c(0)+nu\p^2c(0)+nvcT^m(0)+nw:bc\p c(0): \nonumber \\
&\to&i\left(-\frac{x}{2}+2u-\frac{13 v}{6} -\frac{5 w}{6}\right)
c(1) +i(-2u+3w)\p c(1) \nonumber\\
&&+ 2i(vc T^m(1)+w:bc\p c(1):+u\p^2 c(1).
\end{eqnarray}
The operators inserted at $\hat z=\frac{n-1}{2} \pi\equiv s$ are
acted by BRST operator:
\begin{eqnarray}
&&Q_B\circ\Big\{ \frac{x}{n}c(s)+nu\p^2c(s)+nvcT^m(s)+nw:bc\p
c(s):\Big\} \nonumber \\
&=&\frac{x}{n}c\p
c(s)+n\left(u-\frac{3w}{2}\right) \p c \p^2c(s)+n
\left(u-\frac{13v}{6}+\frac{2w}{3}\right) c\p^3 c(s)\nonumber\\
&&+n(w-v) c\p cT^m(s),
\end{eqnarray}
and then mapped to,
\begin{eqnarray*}
&\to&-i\left(-\frac{x}{2}+2u-\frac{13 v}{6} -\frac{5 w}{6}\right)
c\p c(-1) +i(-2u+3w)\p c\p^2 c(-1)\\
&&+i(-2u+3w)c\p^2 c(-1) -i\left(2u-\frac{13
v}{3}+\frac{4w}{3}\right)c\p^3 c(-1)- 2i(w-v)c\p cT^m(-1).
\end{eqnarray*}
Therefore, $\langle \psi^{(2)}|Q_B|\psi^{(2)}\rangle$ is ready to
be calculated on the unit disk provided values of $x$, $u$, $v$
and $w$ in Table (\ref{solutions}).

The inner product of three wedge states is the $C_{3n-3}$ surface
in $\hat z$ representation. The three punctures are lying at $\hat
z=0\equiv s_1$, $\hat z=\exp((n-1)\pi/2)\equiv s_2$ and $\hat
z=\exp((n-1)\pi)\equiv s_3$, where for notation simplicity, we use
$s_i$, $i=1,2,3$ to denote the locations. The conformal map for
$C_{3n-3}$ to a unit disk is:
\begin{equation}
g(\hat z)=\exp\left(i\frac{4\hat z}{3n-3}\right).
\label{3map}
\end{equation}
Also denote $t_i\equiv g(s_i)$, $i=1,2,3$, the images of $s_i$.
Under the map (\ref{3map}),
\begin{eqnarray}
\psi^{(2)}_{\hat z}(s_i)& \to& \left(\frac{4i}{3}\right) \Big\{
\frac{1}{t_i} \big(-\frac{9 x}{16}+u -\frac{13 v}{12}-\frac{5
w}{12}\big) c(t_i)+\big(-u+\frac{3w}{2} \big)\p c(t_i)\nonumber\\
&& +t_i (v cT^m(t_i) +w :bc\p c(t_i): +u\p^2 c(t_i))\Big\}.
\end{eqnarray}
It is now straightforward to calculate the three point function
$\langle \psi^{(2)}|\psi^{(2)}*\psi
^{(2)}\rangle$, though tedious. In Table (\ref{tableratio}), the
ratios of
\begin{equation}
R=\lim_{n\to\infty}\frac{\langle
\psi^{(0)}|Q_B|\psi^{(0)}\rangle}{\langle
\psi^{(0)}|\psi^{(0)}*\psi^{(0)}\rangle },
\end{equation}
and $\lim_{n\to\infty} {\cal E}/T_{25}$ are present for all the
real valued nontrivial solutions.

\begin{table}[ht]
\caption{The ratios of $R$ and ${\cal E}/T_{25}$ }
\label{tableratio}
\begin{center}
{\renewcommand\arraystretch{1.1}
\begin{tabular}{|c|c|c|c|c|c|}
  \hline
  $x$ & $u$ & $v$ & $w$ & $R$ & ${\cal E}/T_{25}$ \\
  \hline
  $2.62993$ & $-2.60731$ & $0.703291$ & $9.42887$ &$0.641112$&$-2725.74$  \\
  \hline
  $-0.493996$&$0.00603476$&$0.00564685$ & $0.0553046$&$-0.920868$&$-0.81736$\\
  \hline
  $0.12831$ & $0.213274$ & $0.147305$ & $0.112232$&$0.638489$&$ -0.571957$ \\
  \hline
  $0$ & $6.9\times 10^{-13}$ & $4.6\times 10^{-13}$ & $4.6\times
  10^{-13}$ & $ -0.00052$ & $-1.2\times 10^{-36} $\\
  \hline
\end{tabular}}
\end{center}
\end{table}

One can see only the second solution is a real solution for the
theory, different from butterfly case, where two real solutions
were found at next to leading order. For this solution, $R\simeq
-0.921$ is not improved significantly from the leading order
result $-0.912$. But the ratio of $\frac{{\cal E}_C}{T_{25}}\simeq
-0.82$ is improved much compared with the leading order result
$-0.6645$.

\sectiono{Conclusion}
We constructed the leading order solution of Witten's cubic string
field theory based on a \textit{general} twist invariant star
algebra projector using the technique discovered by Okawa
\cite{0311115}. At leading order, the ratio of the kinetic term to
the cubic interaction and the energy density were calculated. We
found that there is a universal relationship between this ratio
and the energy density, independent of the detailed structure of
the projector. From this universal relation, we concluded that for
any twist invariant projector,  the energy density can account for
at most $68.4616\%$ of the the D25-brane tension at the leading
order. To calculate the ratio and energy density, only one
positive number $\lim_{t\to 0}\frac{a_t}{b_t}$ is needed. This
number is defined by two conformal maps:
\begin{itemize}
\item $g_t(\eta)$ in equation (\ref{trans}), which relates the
star product of two regulated projectors to a single regulated
projector. \item $f_1(\hat z)$, which maps the projector after
cutting the local coordinate patch and gluing the left half string
with the right half string to a unit disk with middle point of
string mapped to the origin and middle point of boundary mapped to
$1$.
\end{itemize}
Generically, it is challenging to calculate $g_t(\eta)$. This map
is only known for regulated slivers and butterflies. But from
equation (\ref{General1st}), only one number
$a_t=g_t'(0)^2/g_t(0)$ is needed to write down the leading order
solution. $f_1(\hat z)$ is usually not very hard to figure out. An
example of $f_1(\hat z)$ for generalized butterfly states was
given in equation (\ref{general f1}). At the leading order, when
our results are applied to regulated butterflies, the results of
\cite{0311115} are reproduced. The results based on regulated
slivers were also constructed explicitly. When contracted with the
solution itself, the ratio of the kinetic term to the cubic
interaction is $-0.912356$ and the energy density accounts for
$66.45\%$ of the D25-brane tension.

We also calculated the next to leading order solution based on the
regulated slivers. We found that there was only one real valued
nontrivial solution at this order, different from that in
regulated butterfly situation, where two real solutions were
obtained. The convergence of the solution based on a regulated
slivers is a little bit slower than that based on a regulated
butterflies. The ratio of the kinetic term to the cubic
interaction when contracted with the solution itself is
$-0.920868$, not improved much compared with the leading order
solution. However, the energy density accounts for $81.736\%$ of
the D25-brane tension, much better than the leading order
solution, though still worse than that based on a regulated
butterfly state. It seems that for generalized butterfly states,
the convergence speed of the solutions depend on the generalized
parameter $\alpha$. Since we lack the information about
$g_t(\eta)$ for generalized butterfly states, this conjecture
cannot be proved. An interesting question is that which projector
can produce the most rapidly convergent solutions. Higher order
calculations are of interest to verify the convergence of this
calculation scheme. Sliver state is a good choice since the map
$\tilde f_t(\xi)$ takes a much simpler form than the corresponding
one for butterfly states.

\bigskip
\noindent {\bf Acknowledgements.} The author is especially
grateful to  Y. Okawa  and B. Zwiebach for many discussions and
much advice. Thanks are also due to M. Schnabl for  helpful
discussions in finishing this work. This work was supported by DOE
contract \#DE-FC02-94ER40818.

\newpage
\appendix
\renewcommand{\thesection}{Appendix \Alph{section}.}
\renewcommand{\theequation}{\Alph{section}.\arabic{equation}}
\section{Conformal Transformations of some non-tensor operators}
\setcounter{equation}{0} Most of the operators in the next to
leading order calculations are non-tensor ones. We summarize their
transformation rules under a conformal map $z'=h(z)$:
\begin{eqnarray}
\label{TRANS}
&&h\circ\p c(z)=\p c(z')-\frac{h''(z)}{h'^2(z)}c(z'),\nonumber\\
&&h\circ \p^2 c(z)=h'(z)\p^2c(z')-\frac{h''(z)}{h'(z)}\p
c(z')+\left(2\frac{h''^2(z)}{h'^3(z)}-\frac{h^{(3)}(z)}{h'^2(z)}\right)
c(z'),\nonumber\\
&&h\circ\p^3c(z)=h'^2\p^3
c+\left(3\frac{h''^2}{h'^2}-2\frac{h^{(3)}} {h'}\right) \p c
+\left(
6\frac{h^{(3)}h''}{h'^3}-6\frac{h''^3}{h'^4}-\frac{h^{(4)}}
{h'^2}\right) c,\nonumber\\
&& h\circ (cT^m)=h'cT^m-\left(\frac{13}{4}\frac{h''^2}{h'^3}
-\frac{13}{6}\frac{h^{(3)}}{h'^2}\right) c,\nonumber\\
&& h\circ (:bc\p c:)=h':bc\p c:+\frac{3h''}{2h'}\p c+
\left(\frac{h''^2}{4h'^3}-\frac{2h^{(3)}}{3h'^2} \right) c.
\end{eqnarray}

\section{Some relevant OPEs}
\setcounter{equation}{0} Here we summarize the OPEs needed in the
calculations.
\begin{eqnarray}
\label{OPE} &&c(-t)c(t)=2tc\p c(0)+\frac{1}{3}t^3c\p^3 c(0)-t^3\p
c\p^2 c(0)+{\cal O}(t^4)\nonumber\\
&&c(-t)\p c(t)=c\p c(0)+tc\p^2 c(0)-\frac{3}{2} t^2\p c\p^2 c(0)+
\frac{1}{2}t^2 c\p^3 c(0)+{\cal O}(t^3)\nonumber\\
&& c(-t)\p^2 c(t)=c\p^2 c(0)-t\p c\p^2 c(0)+tc\p^3c(0)+{\cal O}(t^2)\nonumber\\
&&c(-t)c(t)T^m(t)=2tc\p cT^m(0)+{\cal O}(t^2)\nonumber\\
&&c(-t):bc\p c(t):=-\frac{1}{2t}c\p c(0)-\frac{1}{2}c\p^2
c(0)-\frac{t}{4}(\p c\p^2c+c\p^3c)+{\cal O}(t^2)\nonumber\\
&& \p c(-t):bc\p c(t):=-\frac{1}{4t^2} c\p c(0)-\frac{1}{4t}
c\p^2c(0) -\frac{1}{8}(\p c \p^2 c(0)+c\p^3 c(0))+{\cal O}(t)\nonumber\\
&&\p^2 c(-t):bc\p c(t):=-\frac{1}{4t^3}c\p c(0)-\frac{1}{4t^2}
c\p^2 c(0)-\frac{1}{8t}(\p c \p^2 c(0)+c\p^3 c(0))+{\cal O}(1)\nonumber\\
&&:bc\p c(-t)::bc\p c(t):=-\frac{3}{8t^3}c\p c(0)+\frac{15}{16t}
\p c\p^2 c(0)-\frac{7}{48 t} c\p^3 c(0)+{\cal O}(1),\nonumber\\
\end{eqnarray}
where we only keep the necessary orders in $t$.

\end{document}